\documentclass[twocolumn]{aastex63}

\defcitealias{Mikolaitis2018}{M18}
\defcitealias{Mikolaitis2019}{M19}

\newcommand\noallfield{302}
\newcommand\noatmospheric{277}
\newcommand\nobinary{11}
\newcommand\norotat{10}

\received{2019 December 16}
\revised{2020 April 6}
\accepted{2020 April 18}
\submitjournal{ApJS}

\shorttitle{Chemical composition of bright stars in the TESS continuous viewing zone}
\shortauthors{Tautvai\v{s}ien\.{e} et al.}

\begin{document}

\title{CHEMICAL COMPOSITION OF BRIGHT STARS IN THE CONTINUOUS \\ VIEWING ZONE OF THE TESS SPACE MISSION }

\correspondingauthor{Gra\v{z}ina Tautvai\v{s}ien\.{e}}
\email{grazina.tautvaisiene@tfai.vu.lt}

\author{G. Tautvai\v{s}ien\.{e}}
\affiliation{Astronomical Observatory, Institute of Theoretical Physics and Astronomy, 
Vilnius University, \\ Sauletekio av. 3, 10257 Vilnius, Lithuania}

\author{\v{S}. Mikolaitis}
\affiliation{Astronomical Observatory, Institute of Theoretical Physics and Astronomy, 
Vilnius University, \\ Sauletekio av. 3, 10257 Vilnius, Lithuania}

\author{A. Drazdauskas}
\affiliation{Astronomical Observatory, Institute of Theoretical Physics and Astronomy, 
Vilnius University, \\ Sauletekio av. 3, 10257 Vilnius, Lithuania}

\author{E. Stonkut\.{e}}
\affiliation{Astronomical Observatory, Institute of Theoretical Physics and Astronomy, 
Vilnius University, \\ Sauletekio av. 3, 10257 Vilnius, Lithuania}

\author{R. Minkevi\v{c}i\={u}t\.{e}}
\affiliation{Astronomical Observatory, Institute of Theoretical Physics and Astronomy, 
Vilnius University, \\ Sauletekio av. 3, 10257 Vilnius, Lithuania}

\author{H. Kjeldsen}
\affiliation{Astronomical Observatory, Institute of Theoretical Physics and Astronomy, 
Vilnius University, \\ Sauletekio av. 3, 10257 Vilnius, Lithuania}
\affiliation{Stellar Astrophysics Centre, Department of Physics and Astronomy, Aarhus University, \\ Ny Munkegade 120, DK-8000 Aarhus C, Denmark}

\author{K. Brogaard}
\affiliation{Astronomical Observatory, Institute of Theoretical Physics and Astronomy, 
Vilnius University, \\ Sauletekio av. 3, 10257 Vilnius, Lithuania}
\affiliation{Stellar Astrophysics Centre, Department of Physics and Astronomy, Aarhus University, \\ Ny Munkegade 120, DK-8000 Aarhus C, Denmark}

\author{C. von Essen}
\affiliation{Astronomical Observatory, Institute of Theoretical Physics and Astronomy, 
Vilnius University, \\ Sauletekio av. 3, 10257 Vilnius, Lithuania}
\affiliation{Stellar Astrophysics Centre, Department of Physics and Astronomy, Aarhus University, \\ Ny Munkegade 120, DK-8000 Aarhus C, Denmark}

\author{F. Grundahl}
\affiliation{Astronomical Observatory, Institute of Theoretical Physics and Astronomy, 
Vilnius University, \\ Sauletekio av. 3, 10257 Vilnius, Lithuania}
\affiliation{Stellar Astrophysics Centre, Department of Physics and Astronomy, Aarhus University, \\ Ny Munkegade 120, DK-8000 Aarhus C, Denmark}

\author{E. Pak\v{s}tien\.{e}}
\affiliation{Astronomical Observatory, Institute of Theoretical Physics and Astronomy, 
Vilnius University, \\ Sauletekio av. 3, 10257 Vilnius, Lithuania}

\author{V. Bagdonas}
\affiliation{Astronomical Observatory, Institute of Theoretical Physics and Astronomy, 
Vilnius University, \\ Sauletekio av. 3, 10257 Vilnius, Lithuania}

\author{C. Viscasillas V\'azquez}
\affiliation{Astronomical Observatory, Institute of Theoretical Physics and Astronomy, 
Vilnius University, \\ Sauletekio av. 3, 10257 Vilnius, Lithuania}


 
\begin{abstract}

Accurate atmospheric parameters and chemical composition of stars play a vital role in characterizing  physical parameters of exoplanetary systems and understanding of their formation. 
A full asteroseismic characterization of a star is also possible if its main atmospheric parameters are known. 
The NASA Transiting Exoplanet Survey Satellite (TESS) space telescope will play a very important role in searching of exoplanets around bright stars and stellar asteroseismic variability research.  
We have observed all \noallfield ~bright ($V<8$~mag) and cooler than F5 spectral class stars in the northern TESS continuous viewing zone with a 1.65~m telescope at the Mol\.{e}tai Astronomical Observatory of Vilnius University and the high-resolution Vilnius University Echelle Spectrograph. We uniformly determined the main atmospheric parameters, ages, orbital parameters, velocity components, and precise abundances of 24 chemical species 
( C(C$_2$), N(CN), [\ion{O}{1}], \ion{Na}{1}, \ion{Mg}{1}, \ion{Al}{1}, \ion{Si}{1}, \ion{Si}{2}, \ion{Ca}{1}, \ion{Ca}{2}, \ion{Sc}{1}, \ion{Sc}{2}, \ion{Ti}{1}, \ion{Ti}{2}, \ion{V}{1}, \ion{Cr}{1}, \ion{Cr}{2}, \ion{Mn}{1}, \ion{Fe}{1}, \ion{Fe}{2}, \ion{Co}{1}, \ion{Ni}{1}, \ion{Cu}{1}, and \ion{Zn}{1})  for  \noatmospheric ~slowly rotating single stars in the field. About 83\,\% of the sample stars exhibit the Mg/Si ratios greater than 1.0 and may potentially harbor rocky planets in their systems. 

\end{abstract}



\section{Introduction} \label{sec:intro}


The NASA Transiting Exoplanet Survey Satellite (TESS) is an ongoing space mission with primary goal to search for planets in systems of bright and nearby stars \citep{Ricker2015}. 
The two-year mission is planned to observe southern and northern ecliptic hemispheres, each divided into 13 sectors. A total of 26 sectors will cover 85\,\% of the sky. 
Those sectors in each hemisphere overlap at the ecliptic poles, forming circular regions -- continuous viewing zones (CVZs). The CVZs will have an almost full year observing coverage; such an advantage would be significant in finding planets of longer orbital periods.
The TESS mission at a 30 minute cadence will observe all stars in its field of view, yet for more than 200,000 stars selected for a Candidate Target List (CTL), a two minute cadence will be used. A complex mechanism of CTL compilation is described in more detail in \citet{Stassun18}. 

TESS was launched on 2018 April and has already finished its first year observations of the southern ecliptic hemisphere and now is pointing to sectors of the northern ecliptic hemisphere. As of 2019 December, TESS mission results are 37 confirmed planets and 1417 planet candidates \citep[NASA exoplanet archive;][]{Akeson13}. However, the former number will likely be increased, since \citet{Barclay18} estimated that $1250\pm{70}$ exoplanets will be found using the two minute cadence mode and 3100 and 10,000 exoplanets -- using the 30 minute cadence mode around bright dwarfs and fainter stars, respectively.

Since objects of the TESS mission are much nearer and 10--100 times brighter than those of the Kepler mission \citep{Borucki2010,Koch2010}, they are excellent targets for ground-based observations \citep{Ricker2015, Barclay18}. Data from ground-based observations are a crucial ingredient in order to characterize identified exoplanets and especially their atmospheres. 
The chemical composition of the protoplanetary disk and planet formation pathways are linked to the bulk composition of the parent star. Furthermore, different stellar Galactic subcomponents could produce planets with different properties. Thus, it becomes crucial to determine the bulk chemical composition of the stars in different Galactic populations. The homogeneously determined detailed stellar abundances from ground-based observations are starting to be transferred to the models that help to deduce properties of planet building blocks and exoplanets themselves \citep[see e.g.,][]{Santos17,Cabral19, Bitsch20}. Moreover, TESS is expected to deliver major improvements in characterizing planets. Different from the Kepler mission \citep{Thompson2018}, due to the large signal-to-noise ratio (S/N) of TESS planet candidates, the atmospheric characterization of Neptune and Earth-sized planets will be feasible with current and future technology and amenable for some hundreds of them \citep{Kempton2018}.

Since only around one third of bright ($V<8$~mag{\bf )} stars have high-resolution spectroscopic studies, in 2016, we started a Spectroscopic and Photometric Survey of bright stars in the northern hemisphere. In papers by \citet[hereafter \citetalias{Mikolaitis2018}]{Mikolaitis2018} and \citet[hereafter \citetalias{Mikolaitis2019}]{Mikolaitis2019}, we published atmospheric parameters and detailed chemical compositions for 249 bright dwarf stars located in two preliminary ESA PLATO fields: STEP02 and NPF 
(\citealt{Rauer2014,Rauer2016}). With the current work, we aimed to observe high-resolution spectra for all both dwarf and giant stars with $V<8$~mag and cooler than F\,5 spectral type in the TESS northern CVZ and to determine homogeneously their main parameters and detailed chemical compositions.  We hope our work will be useful in characterizing exoplanets potentially discovered by TESS around those stars and for the asteroseismic stellar analyses.

\section{Observations and method of analysis} \label{sec:obs-methods}

\begin{figure}
\epsscale{1.15}
\plotone{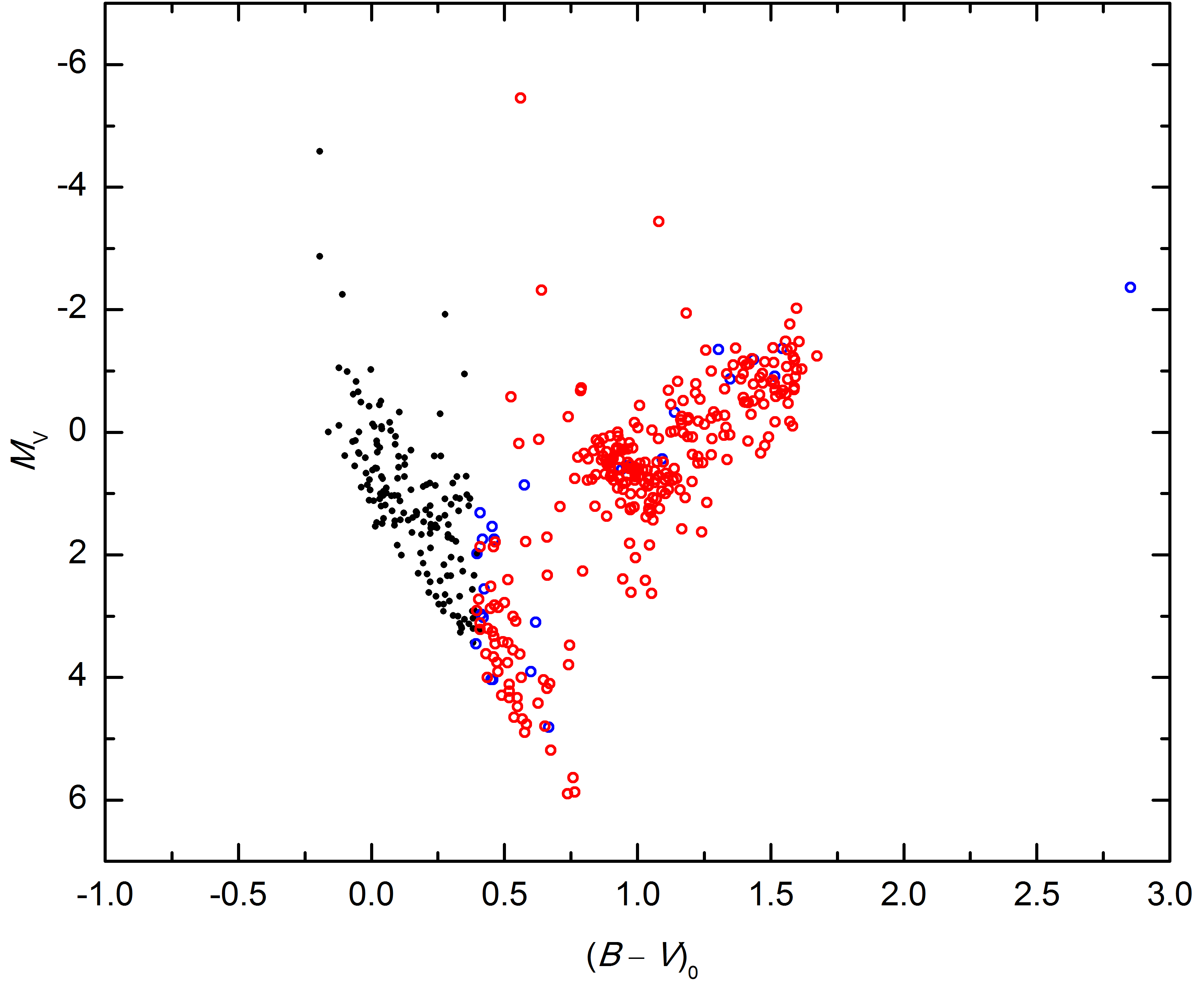}
\caption{Color–magnitude diagram of stars in the investigated TESS CVZ field. The FGK spectral type stars investigated in this work are marked as open circles; objects with determined parameters are indicated as red circles and the peculiar stars as blue circles.
\label{fig:HR1}}
\end{figure}

\begin{figure}
\epsscale{1.15}
\plotone{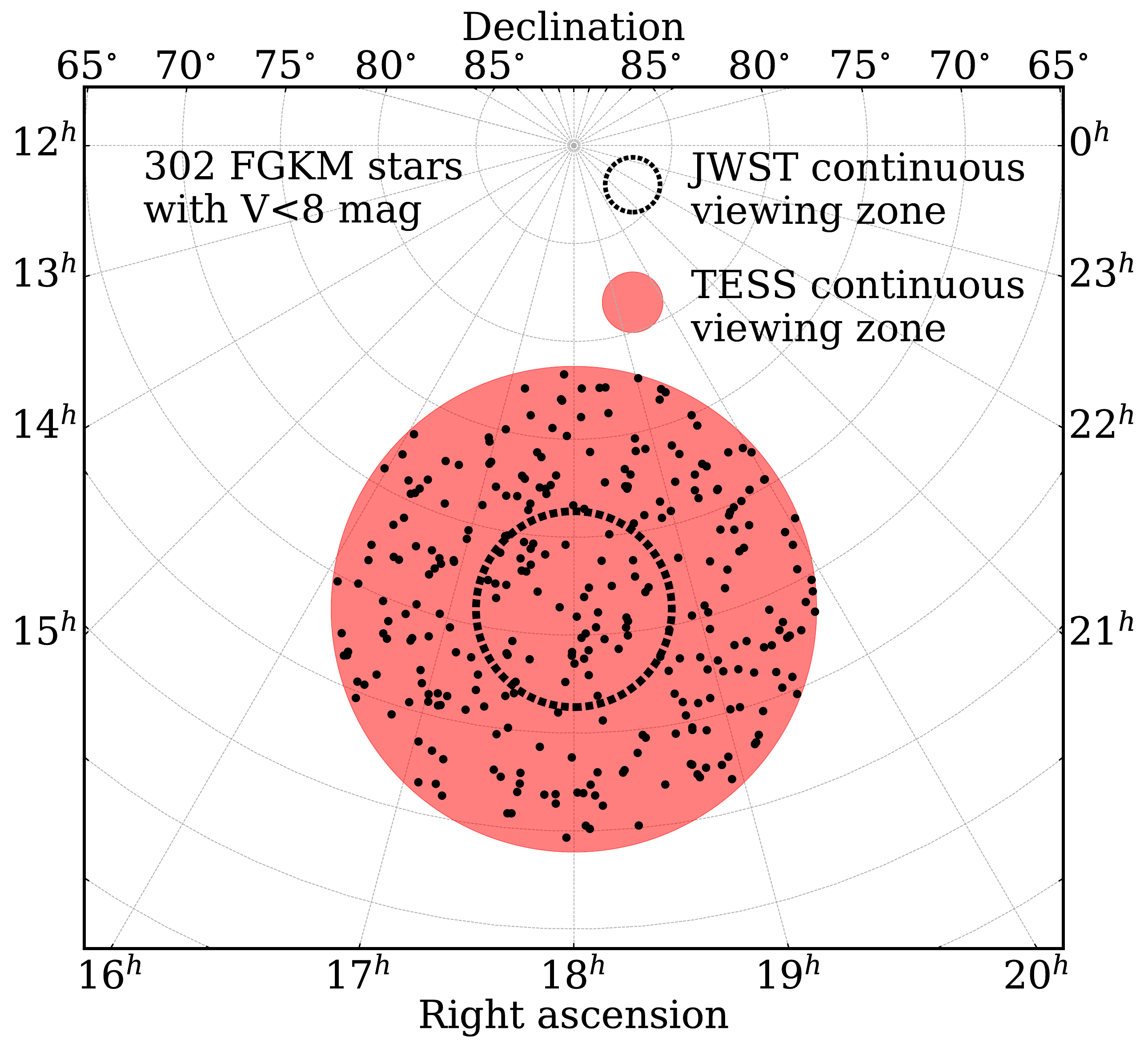}
\caption{Positions (R.A. and decl. in hours and degrees, respectively) of the program stars (black dots). The CVZs of TESS and JWST are marked as well.
\label{fig:CVZ}}
\end{figure}

\subsection{Target Selection and Observations}\label{subsec:targets}

Having our observational possibilities and methods of analysis developed in mind, we decided to observe high-resolution spectra of all bright ($V<8$~mag) F5 and cooler stars ($T_{\rm eff}$~$<$~6500~K) in the whole TESS northern CVZ  which is a region around the northern ecliptic pole with a diameter of around 24 degrees. 

We examined a color-magnitude diagram of all $V~<~8$~mag and $(B-V)>0.39$~mag stars in the selected field (Figure~\ref{fig:HR1}). In \citetalias{Mikolaitis2018}, we showed that $T_{\rm eff}$~$<$~6500~K corresponds to approximately $(B-V)>0.39$~mag. 
In this way, we found \noallfield ~stars in the selected field that met these criteria (see Figure~\ref{fig:CVZ}) and we have observed all of them during the period of 2018--2019. Of the observed stars, 53 fall within the 5 degree radius CVZ of the upcoming James Webb Space Telescope (JWST) mission \citep{Gardner06}. 

We used the 1.65~m telescope at the Moletai Astronomical Observatory of Vilnius University in Lithuania that is equipped with the high-resolution Vilnius University Echelle Spectrograph (VUES; \citealt{Jurgenson2016}). This spectrograph has a wavelength coverage from 400 to 900~nm in $R\sim$36,000, $\sim$51,000, and $\sim$68,000 resolution modes. For our work, we used the $\sim 68,000$ mode for the M spectral type stars and the $\sim 36,000$ mode for other objects. Exposure times varied between 900 and 2400~s and S/Ns varied between 75 and 200 with the median value at 96, depending on stellar magnitudes.  The VUES data reduction was accomplished on the site using an automated pipeline described by \citet{Jurgenson2016}.

\subsection{Radial Velocity Determination and Identification of
Double-line Binaries and Fast-rotating Stars} \label{subsec:kinematic}

\begin{figure}
\epsscale{1.2}
\plotone{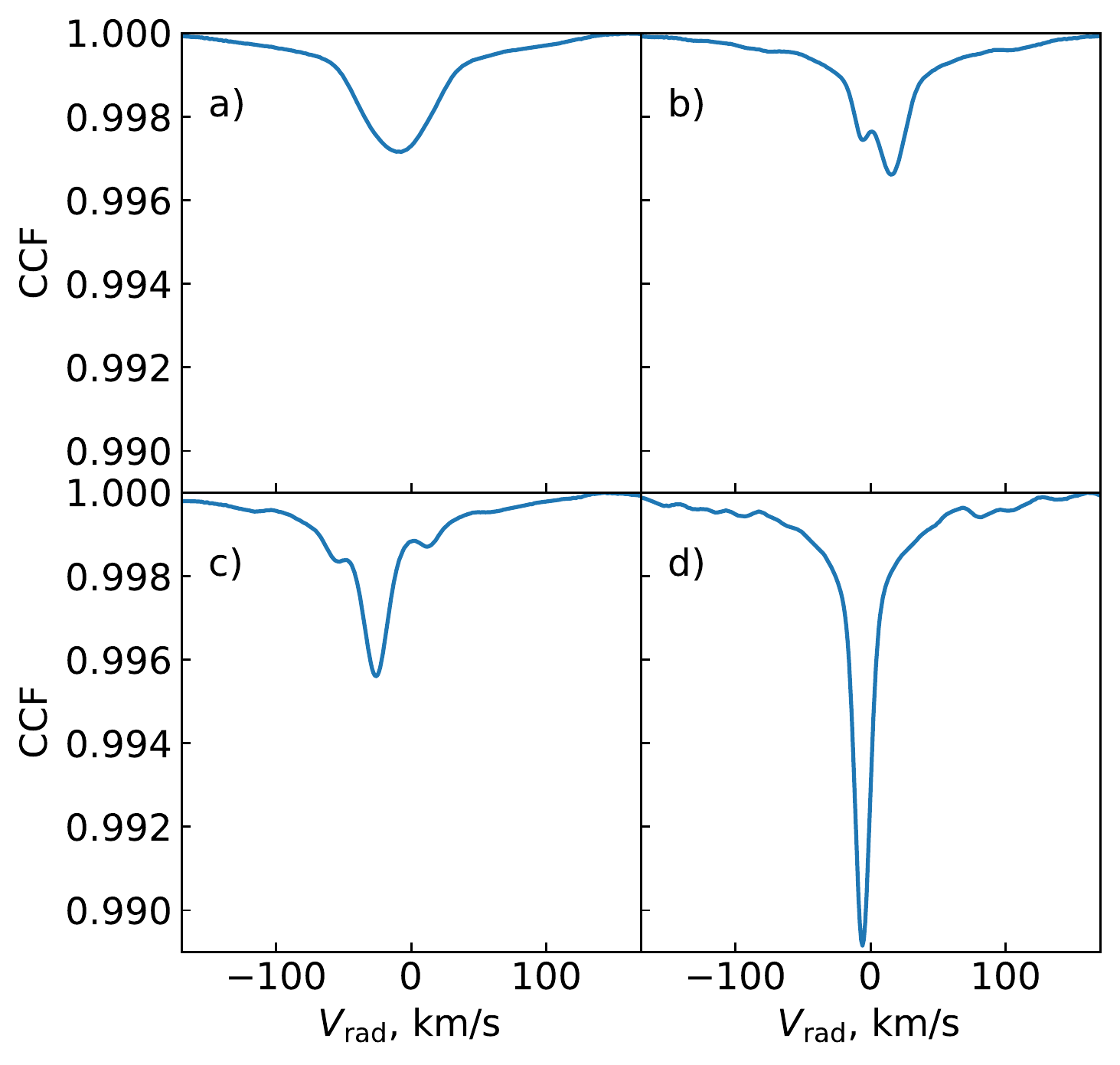}
\caption{Examples of CCFs produced for calculating the radial velocities and detection of double-line binary stars: (a) the fast-rotating star HD~155513, (b) the double-line spectroscopic binary HD~180160, (c) HD~160780 showing three profiles, and (d) a typical slow-rotating star HD~149843.}  
\label{fig:ccf}
\end{figure}

For an initial spectral analysis, we used 
the standard cross-correlation function (CCF) method to obtain spectroscopic radial velocity values. 
The CCF revealed \nobinary~double-line binaries  and \norotat~fast-rotating stars, which we postponed for a further analysis. Figure~\ref{fig:ccf} shows the CCF examples. 
It was not possible to measure equivalent widths of lines for those fast-rotating stars with a satisfactory quality because of the broad and blended lines. The classical equivalent width method that we used to determine the main stellar atmospheric parameters could not be applied to the fast-rotating stars with strongly broadened and diminished lines. From the subsequent analysis, we also excluded the four coolest stars (see Section~\ref{subsec:parameters}) with severe line-blending.
In this way, \noatmospheric~stars filled our final sample of the present analysis.

\subsection{Stellar Atmospheric Parameters and Chemical Composition} \label{subsec:parameters}

In order to determine the main stellar atmospheric parameters (effective temperature, $T_{\rm eff}$; surface gravity, ${\rm log}~g$; microturbulence velocity, $v_{\rm t}$; and metallicity $\langle{\rm{[Fe/H]}}\rangle$), we adopted a classical method of  equivalent widths of atomic neutral and ionized iron lines. 
We used a combination  of the DAOSPEC (\citealt{Stetson2008}) and MOOG (\citealt{Sneden1973}) codes the same way as the Vilnius node was using in the $Gaia$-ESO Survey (see \citealt{Smiljanic2014} and \citetalias{Mikolaitis2018}).

Detailed abundances of 24 chemical species were determined applying a spectral synthesis method with the TURBOSPECTRUM code (\citealt{Alvarez1998}). The spectral analysis was done using a grid of MARCS stellar atmosphere models \citep{Gustafsson2008} and the Solar abundances by \citet{Grevesse07}. 
Atomic lines were selected from the $Gaia$-ESO line-list by \citet{Heiter2015}. We have also used the molecular line lists: $\rm{C}_{2}$~\citep[][]{Brooke2013,Ram2014}; CN~\citep[][]{Sneden2014}; CH~\citep[][]{Masseron2014}; SiH~\citep[][]{Kurucz1993}; FeH~\citep[][]{Dulick2003}; CaH~(B. Plez, private communication); and OH, MgH, NH~(T. Masseron, private communication). 
For the carbon abundance determination, we used two regions: the ${\rm C}_2$ Swan (1, 0) band head at 5135~{\AA} and the ${\rm C}_2$ Swan (0, 1) band head at 5635~{\AA}. For the nitrogen abundance determination, we used $\mathrm{^{12}C^{14}N}$ molecular lines in the region 7980--8005~\AA. 
The oxygen abundance was determined from the forbidden [O\,{\sc i}] line at 6300~{\AA}. These elements require a more detailed analysis, as they are bound by the molecular equilibrium. 
First, we performed a couple of iterations until the determinations of carbon and oxygen abundances converged. After that, we used both carbon and oxygen values to determine the abundance of nitrogen.

For more details of the chemical composition analysis, we refer to our recent   studies (\citetalias{Mikolaitis2019}; \citealt{Stonkute2020}).

The applied method did not work for four M3 and cooler stars: V*~HW~Dra, V*~GP~Dra, and V*~TX~Dra are M3--M5 type stars with $Gaia$ DR2 $T_{\rm eff}\approx$3300~K; a specific analysis is also needed for V*~UX~Dra which is classified as a C-N5 carbon star with $T_{\rm eff}\approx 2817$~K (\citealt{Lambert1986}).

\subsection{Stellar Ages}
\label{subsec:ages}

To calculate the stellar ages, and their errors, we used the code UniDAM (the unified tool to estimate distances, ages and masses) by \citet{Mints2017}. The code uses a Bayesian approach and the PARSEC isochrones \citep{Bressan2012}. As an input, we used the stellar atmospheric parameters determined in this work together with the \textit{J}, \textit{H}, and \textit{K} magnitudes from the the Two Micron All-Sky Survey (2MASS) \citep{Skrutskie2006} and the $W1$ and $W2$ magnitudes from AllWISE \citep{Cutri14}. 

Crossmatching of our sample stars with the 2MASS and AllWISE catalogs provided us data for 273 stars that had entries in at least one infrared photometry study: 198 stars had magnitudes from both, and 75 had magnitudes from 2MASS. After calculating the ages, we discarded 59 stars that had flags reported after calculation that meant either unreliable photometry, or the result being off the model grid, or just an unreliable determination (see Section 6.1 in \citealt{Mints2017} for more interpretations). Finally, we were left with 214 stars for which we report the derived ages in this work.

\subsection{Kinematic Properties}
\label{subsec:kinematics}

The main kinematic parameters ($R\mathrm{_{mean}}$, $z\mathrm{_{max}}$, $e$, \textit{U, V, and W}) for the stars were calculated using the python-based package for galactic-dynamics calculations \textit{galpy}\footnote{http://github.com/jobovy/galpy} by \citet{Bovy15}. The parallaxes, proper motions, and coordinates required for \textit{galpy} were taken from the {\it Gaia} data release 2 (DR2) catalog \citep{Luri2018, Katz2019}. 

The \textit{galpy} was set to integrate orbits for 5~Gyr. Observational errors were estimated using 1000~Monte Carlo calculations according to the errors in the input parameters. The position and movement of the Sun are those from \citet{Bovy2012} ($R_{\rm gc\odot}=8$~kpc and $V_{\odot}=220$~km\,s$^{-1}$), the distance from the Galactic plane $z_{\odot}=0.02$~kpc \citep{Joshi07}, and the LSR from \citet{Schonrich10} (\textit{U, V, W} = 11.1, 12.24, 7.25~km\,s$^{-1}$).

\subsection{Errors on Atmospheric Parameters and Abundances}
\label{sec:erroratmospheres}

The errors on the atmospheric parameters were estimated as follows: 

\begin{table}
\caption{Errors Due to the Uncertain Continuum Placement and Equivalent width Measurement, Based on the Monte Carlo Simulations.}
\centering
\begin{tabular}{lrrr}
\hline\hline
 & S/N=25 & S/N=50 & S/N=75 \\
\hline
\multicolumn{4}{c}{TYC~3910-1710-1}\\
\multicolumn{4}{c}{$T_{\rm eff}=4458$\,K, ${\rm log}~g=2.72$, ${\rm [Fe/H]}=-0.14$}\\

$\sigma_{T_{\rm eff}}$	&  	49	&   	45	&   	32	\\
$\sigma_{{\rm log}~g}$ 	&	0.09	&	0.09	&	0.09	\\
$\sigma_{{\rm [Fe/H]}}$  &	0.03	&	0.03	&	0.01	\\
$\sigma_{v_{\rm t}}$  	&	0.08	&	0.08	&	0.04	\\
\hline
\ion{Na}{1} 	&	0.08	&	0.08	&	0.05	\\
\ion{Mg}{1} 	&	0.10	&	0.06	&	0.03	\\
\ion{Al}{1} 	&	0.09	&	0.07	&	0.04	\\
\ion{Si}{1} 	&	0.05	&	0.04	&	0.03	\\
\ion{Si}{2} 	&	0.08	&	0.08	&	0.05	\\
\ion{Ca}{1} 	&	0.10	&	0.07	&   0.05	\\
\ion{Ca}{2} 	&	0.09	&	0.08	&	0.07	\\
\ion{Sc}{1} 	&	0.12	&	0.08	&	0.06	\\
\ion{Sc}{2} 	&	0.11	&	0.11	&	0.06	\\
\ion{Ti}{1} 	&	0.10	&	0.08	&	0.06	\\
\ion{Ti}{2} 	&	0.11	&	0.09	&	0.03	\\
\ion{V}{1} 	   &	0.09	&	0.08	&	0.06	\\
\ion{Cr}{1} 	&	0.09	&	0.09	&	0.05	\\
\ion{Cr}{2} 	&	0.12	&	0.09	&	0.05	\\
\ion{Mn}{1} 	&	0.10	&	0.08	&	0.06	\\
\ion{Fe}{1} 	&	0.09	&	0.06	&	0.04	\\
\ion{Fe}{2} 	&	0.13	&	0.09	&	0.07	\\
\ion{Co}{1} 	&	0.09	&	0.07	&	0.01	\\
\ion{Ni}{1} 	&	0.06	&	0.05	&	0.02	\\
\ion{Cu}{1} 	&	0.07	&	0.06	&	0.06	\\
\ion{Zn}{1} 	&	0.12	&	0.09	&	0.06	\\
\hline

 \label{tab:montecarlo}
\end{tabular}

\end{table}

\begin{table*}
\caption{Median Effects on the Derived Abundances Resulting from the Atmospheric Parameter Uncertainties for the Sample Stars.
}
\label{tab:sensitivity}
      \[
         \begin{tabular}{lcrrcrrcc}
            \hline
            \noalign{\smallskip}
	    El & 
	    ${ \Delta T_{\rm eff} }$ & 
            ${ \Delta \log g }$ & 
            $\Delta {\rm [Fe/H]}$ & 
            ${ \Delta v_{\rm t} }$ 
            & $\sigma_{\rm{scat}}^{a}$  
     	     & $N_{max}^{b}$ 
     	     & $ \sigma_{\rm total\left[\frac{El}{H}\right]}^{c} $  
	     & $ \sigma_{\rm all\left[\frac{El}{H}\right]}^{d} $  
	     \\ 
	     & K & & & km\,s$^{-1}$ & & & & \\
            \noalign{\smallskip}
            \hline
            \noalign{\smallskip}
C (C$_{2}$) &  0.00 & 0.03 & 0.01 & 0.01 & 0.02 & 2 &  0.03  &  0.04  \\
N (CN) & 0.06 & 0.05 & 0.01 & 0.01 & 0.05 & 7 & 0.08 &  0.08  \\
O ([O\,{\sc i}]) & 0.01  & 0.08 & 0.07 & 0.00 & 0.06 & 1  & 0.11    &  0.13\\   
\ion{Na}{1} 	&	0.00	&	0.03	&	0.02	&		0.02	&	0.03	&	4	&	0.05	&			0.05	\\
\ion{Mg}{1} 	&	0.01	&	0.07	&	0.02	&		0.03	&	0.06	&	5	&	0.08	&			0.12	\\
\ion{Al}{1} 	&	0.01	&	0.02	&	0.03	&		0.03	&	0.04	&	5	&	0.03	&			0.07	\\
\ion{Si}{1} 	&	0.01	&	0.02	&	0.02	&		0.02	&	0.03	&	14	&	0.02	&			0.03	\\
\ion{Si}{2} 	&	0.01	&	0.05	&	0.01	&		0.02	&	0.07	&	7	&	0.06	&			0.10	\\
\ion{Ca}{1} 	&	0.02	&	0.07	&	0.02	&		0.02	&	0.06	&	31	&	0.08	&			0.10	\\
\ion{Ca}{2} 	&	0.03	&	0.06	&	0.02	&		0.04	&	0.05	&	7	&	0.07	&			0.09	\\
\ion{Sc}{1} 	&	0.04	&	0.03	&	0.02	&		0.03	&	0.08	&	7	&	0.05	&			0.11	\\
\ion{Sc}{2} 	&	0.02	&	0.08	&	0.02	&		0.04	&	0.03	&	12	&	0.10	&			0.10	\\
\ion{Ti}{1} 	&	0.04	&	0.04	&	0.01	&		0.02	&	0.03	&	81	&	0.07	&			0.08	\\
\ion{Ti}{2} 	&	0.01	&	0.08	&	0.04	&		0.04	&	0.04	&	19	&	0.09	&			0.09	\\
\ion{V}{1}  	&	0.03	&	0.01	&	0.02	&		0.04	&	0.05	&	8	&	0.05	&			0.07	\\
\ion{Cr}{1} 	&	0.02	&	0.03	&	0.02	&		0.03	&	0.05	&	21	&	0.05	&			0.06	\\
\ion{Cr}{2} 	&	0.02	&	0.08	&	0.02	&		0.04	&	0.03	&	2	&	0.10	&			0.11	\\
\ion{Mn}{1} 	&	0.03	&	0.03	&	0.02	&		0.04	&	0.05	&	14	&	0.05	&			0.07	\\
\ion{Fe}{1} 	&	0.02	&	0.03	&	0.02	&		0.03	&	0.05	&	138	&	0.05	&			0.06	\\
\ion{Fe}{2} 	&	0.01	&	0.08	&	0.03	&		0.04	&	0.04	&	11	&	0.09	&			0.12	\\
\ion{Co}{1} 	&	0.01	&	0.01	&	0.01	&		0.02	&	0.05	&	7	&	0.02	&			0.07	\\
\ion{Ni}{1} 	&	0.01	&	0.02	&	0.01	&		0.04	&	0.03	&	30	&	0.03	&			0.07	\\
\ion{Cu}{1} 	&	0.03	&	0.02	&	0.02	&		0.02	&	0.05	&	6	&	0.04	&			0.06	\\
\ion{Zn}{1} 	&	0.01	&	0.02	&	0.02	&		0.02	&	0.09	&	3	&	0.04	&			0.11	\\

\hline
         \end{tabular} 
      \]

Notes.\\
$^{a} \sigma_{\rm{scat}}$ stands for the median line-to-line scatter.\\ 
$^{b}{\rm N}_{max}$ presents the number of lines investigated.\\
$^{3c}\sigma_{\rm total([El/H])} $ stands for the median of the quadratic sum  of all four effects on [El/H].\\
$^{d}\sigma_{\rm all([El/H])} $ is a median of the combined effect of $ \sigma_{\rm total([El/H])} $ and the line-to-line scatter $\sigma_{\rm{scat}}$. 

   \end{table*}

\begin{itemize}
\item
We tested the sensitivity of computed atmospheric parameters to the quality of the spectra. As a representative star for this test, we choose TYC\,3910-1710-1 -- a giant star with high signal-to-noise spectra. We artificially degraded the spectra of this star to S/Ns of 25, 50, and 75 per pixel and determined its atmospheric parameters. Of the randomly degraded spectra for each S/N, 100 were used and the generated statistics are presented in Table~\ref{tab:montecarlo}. The same test for dwarf stars was shown in \citetalias{Mikolaitis2018}.
\item
The uncertainties in the atomic parameters of the used lines create a scatter of measured iron abundances and also an error in a linear regression fit that can be directly propagated to follow the uncertainties of atmospheric parameters. Therefore, the uncertainties for each of the main atmospheric parameters are provided for every star in machine-readable Table~\ref{tab:CDS} and they are computed the same way as published in the description of the $Gaia$-ESO Vilnius node by \citet{Smiljanic2014}.
The median errors measured by the algorithm in the full stellar sample are $\sigma_{T_{\rm eff}}$=60~K, $\sigma_{{\rm log}~g}$=0.21~dex, $\sigma_{{\rm [Fe/H]}}$=0.11~dex, and $\sigma_{v_{\rm t}}$=0.23~${\rm km~s}^{-1}$.
\end{itemize}

We have made several tests to understand the error budget in our chemical abundance measurements:
\begin{itemize}
\item 
We used the same 300 generated spectra of the giant star TYC\,3910-1710-1 for three S/N values to measure abundances in order to estimate their sensitivity to quality of the spectrum. These results are provided in Table~\ref{tab:montecarlo}. The same test for dwarf stars was accomplished in \citetalias{Mikolaitis2019}.
\item
Evaluation of the line-to-line scatter is a way to estimate random errors if the number of lines is large enough. A median of the standard deviation for a given element, $\sigma_{\rm{scat}}^{*}$, is presented in the sixth column of Table~\ref{tab:sensitivity}.

\item\begin{figure*}
\begin{minipage}{0.3\linewidth}
\includegraphics[width=\textwidth]{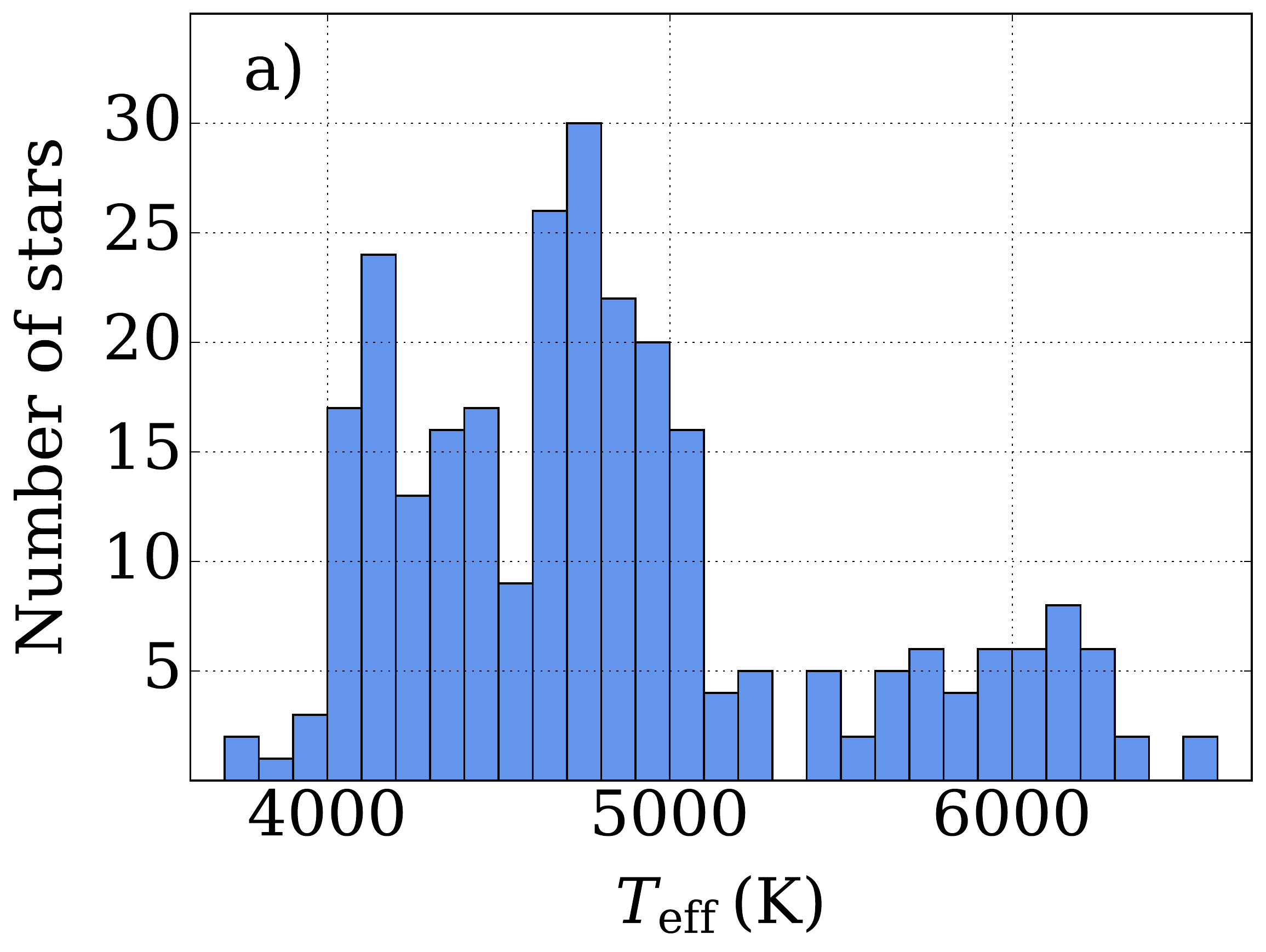}
\end{minipage}%
\hfill
\begin{minipage}{0.3\linewidth}
\includegraphics[width=\textwidth]{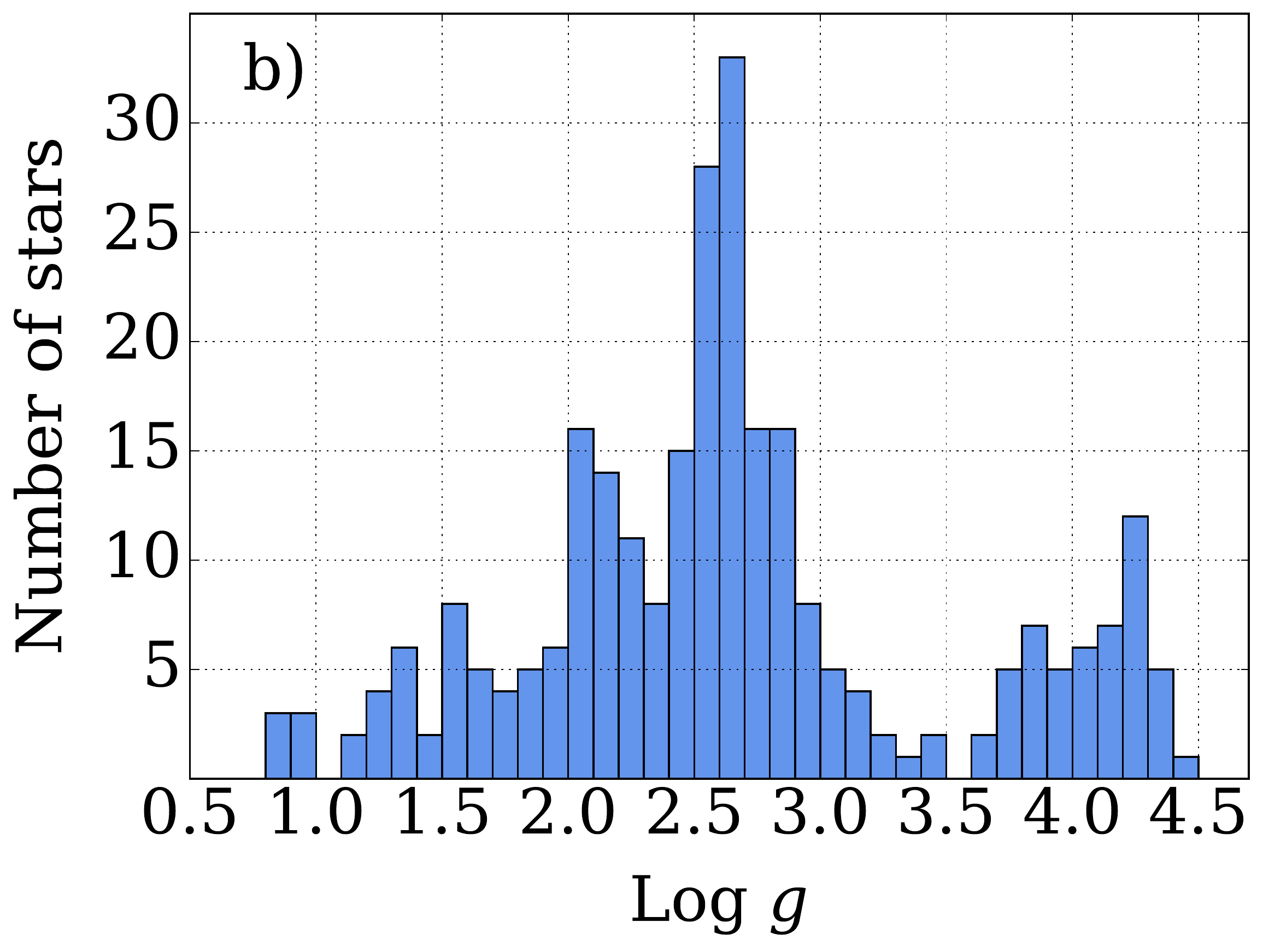}
\end{minipage}%
\hfill
\begin{minipage}{0.3\linewidth}
\includegraphics[width=\textwidth]{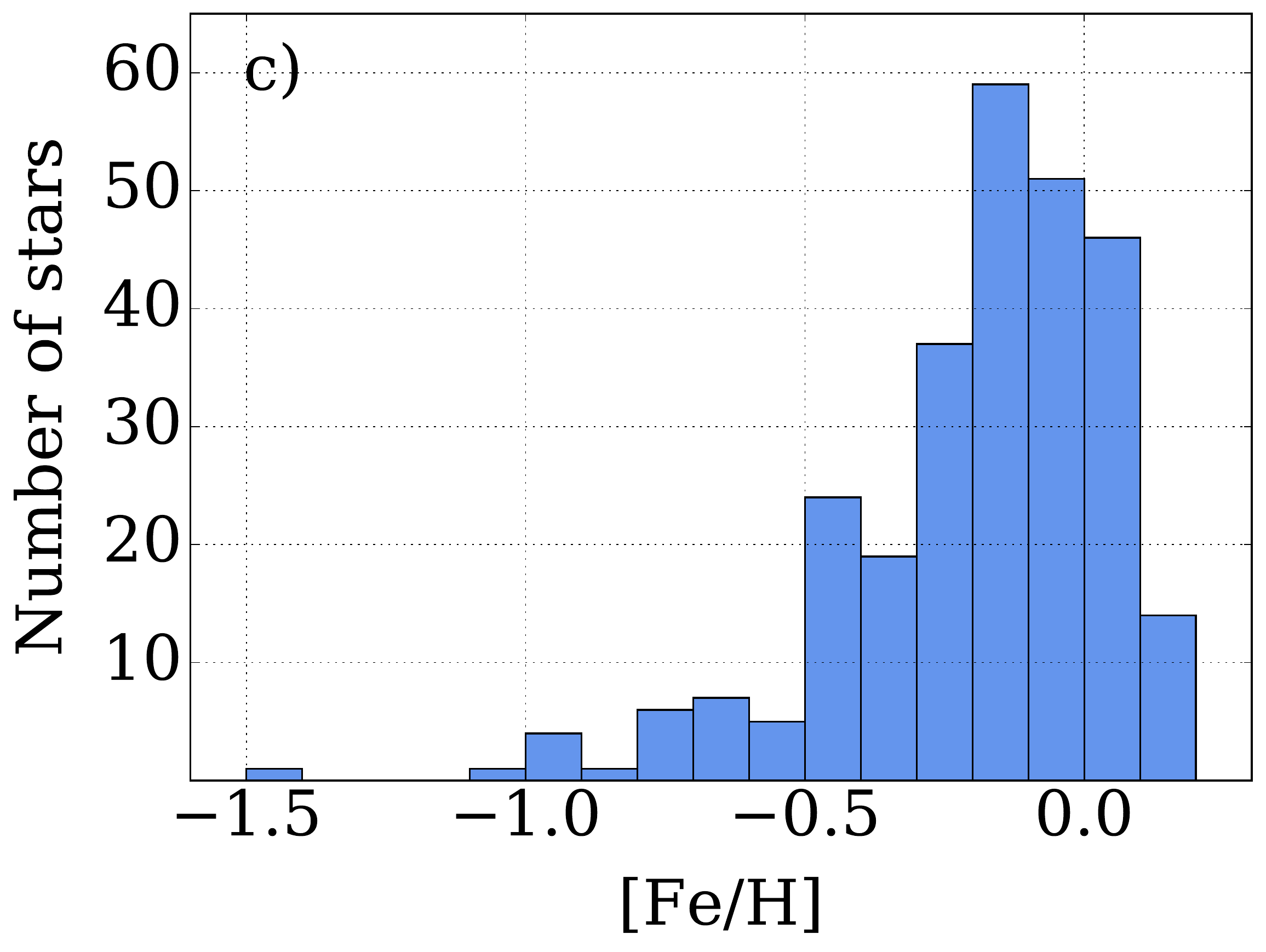}
\end{minipage}
\caption{Histograms of the determined spectroscopic parameters ($T_{\rm eff}$, log $g$, and [Fe/H]) for the sample stars.}
\label{fig:histograms}
\end{figure*}

The uncertainties of the main atmospheric parameters were propagated into the errors of chemical abundances. The median errors of this type over the stellar sample are provided in Table~\ref{tab:sensitivity}.                         
\end{itemize}

The final error for every element for every star that is given in machine-readable Table~\ref{tab:CDS} is a quadratic sum of effects due to uncertainty in four atmospheric parameters and the abundance scatter given by the lines.

In Section~3.6 of \citetalias{Mikolaitis2019}, we have discussed that it should be safe to use the classical LTE approach to compute elemental abundances in the metallicity regime of our sample stars. The NLTE effects for the majority of elemental abundances in our sample are negligible; the NLTE corrections were adopted just for the potassium abundances as they were determined from the large 7698.9~\AA ~line. For manganese and copper, we have accounted for a hyperfine splitting as described in \citetalias{Mikolaitis2019}. 

Since abundances of C, N, and O are bound together by the molecular equilibrium in the stellar atmospheres, we investigated how an error in one of them typically influences the abundance determination of another. We determined that $\Delta$[O/H] = 0.10 causes $\Delta$[C/H] = 0.02 and $\Delta$[N/H] = 0.04, and $\Delta$[C/H] = 0.10 causes $\Delta$[N/H] = −0.11 and $\Delta$[O/H] = 0.02, while $\Delta$[N/H] = 0.10 has no effect on either the carbon or the oxygen abundances.

\begin{figure}
 \includegraphics[width=\columnwidth]{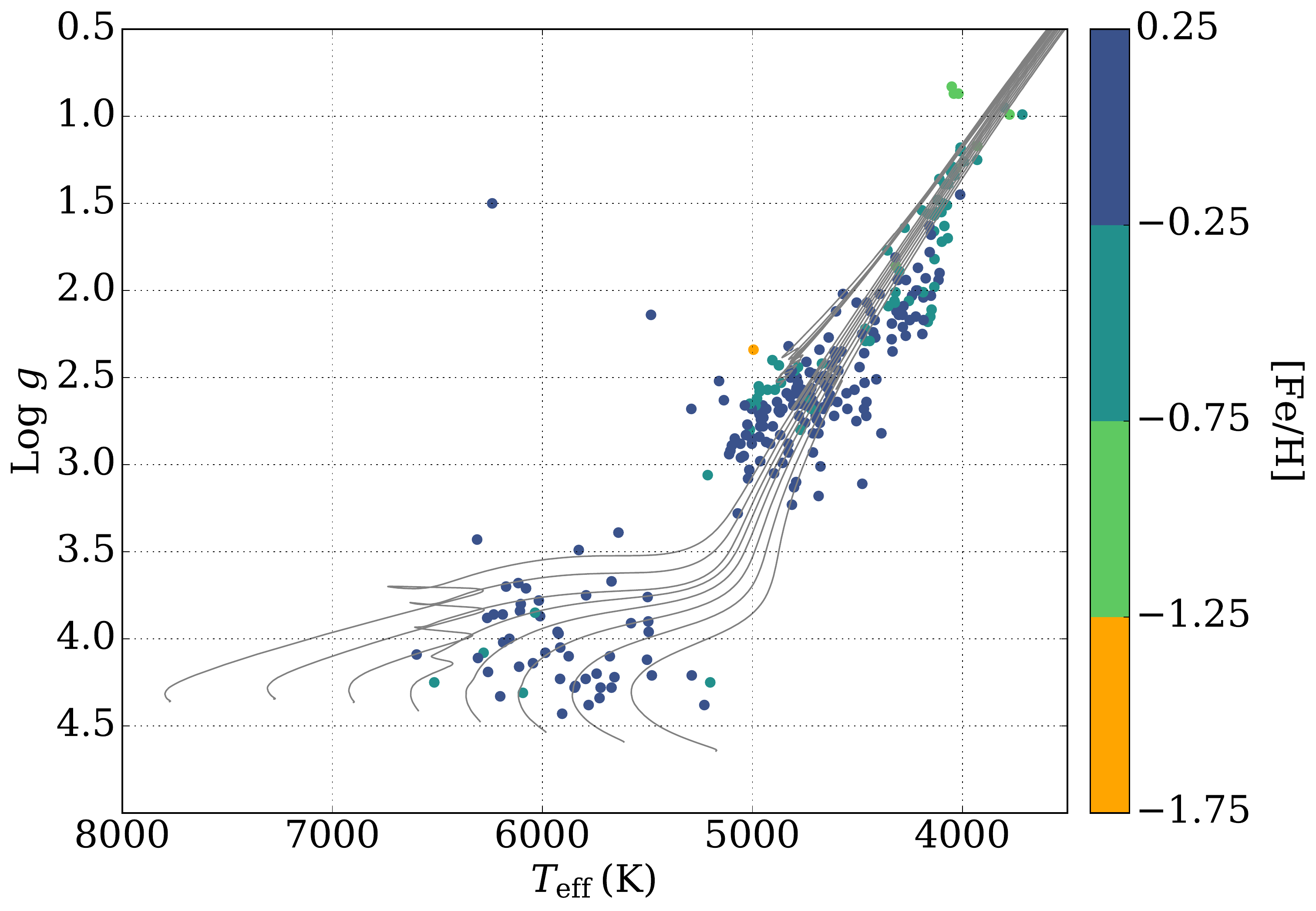}
\caption{Log $g$ and $T_{\rm eff}$ diagram of the investigated stars (dots) with
metallicity coded by color. Evolutionary sequences by \citet{Bressan2012} with
masses between 0.8 and 1.5\,$M_{\odot}$ and $Z_{\rm ini}=0.01$ are plotted as gray solid
lines.}
\label{fig:HR2}
\end{figure}

\section{Results and discussion}\label{sec:results}

The initial CCF analysis of spectra revealed \nobinary ~double- or even triple-line stellar systems.   
Of the 11, four are known spectroscopic binaries: HD\,155902, HD\,165700, $\chi$~Dra, and HD\,180160 (\citealt{Nordstrom2004}). The newly detected double-line spectroscopic binaries are the following stars:  HD~152274, HD~145222, HD~170527, HD~184756, and V*~AZ~Dra, while the triple-line spectrum was found for HD~165988 and HD~160780.
Of the stars, 10 appeared to be quite fast rotating ($V_{rot}\ge 20$~km\,s$^{-1}$:  HD~161128, HD~164983, HD~179729, HD~171044, HD~238865, HD~173605, V*~omi~Dra, HD~155513, HD~164330, and HD~165522. These stars were postponed for a further  investigation requiring different methods of analysis and additional photometric and spectral observations.
And finally, four stars had too low effective temperatures for our method of analysis and also will be investigated later.
Thus, a full characterization was performed for a sample of the remaining  \noatmospheric~stars.

\subsection{Atmospheric Parameters and Elemental Abundances}

Our sample of \noatmospheric~slowly rotating stars have 
$T_{\rm eff}$ between 3700 and 6600~K with a peak at 4700~K, [Fe/H] are from $-1.5$~dex to 0.25~dex with a peak at $-0.25$~dex, and log\,$g$ are from 0.8 to 3.5 with a peak at 2.7 for giants and from 3.6 to 4.5 with a peak at 4.3 for dwarfs (more detailed distributions are presented in Figure~\ref{fig:histograms}). A log\,$g$ versus $T_{\rm eff}$ diagram of the stars colored according to their metallicity is presented in Figure~\ref{fig:HR2}.

From the $fe\_h$ catalog of the SIMBAD (\citealt{Wenger2000}) database and the PASTEL catalog (\citealt{Soubiran2010}) we found that there are only few studies that have derived stellar parameters for some stars of our sample.
We have only 47 stars that were observed before with atmospheric parameters delivered using high-resolution spectrographs. This spectroscopic comparison sample is collected from \citet{1990ApJS...74.1075M, 2007A&A...475.1003H, 2007AJ....133.2464L, 2011A&A...526A..71D, 2011AJ....141...90L, 2011MNRAS.415.1138P, 2012A&A...542A.104B, 2012AJ....144...20A, 2014AJ....147..136R, 2014ApJ...785...94L, 2014ApJ...791...58A, 2015A&A...580A..24D, 2016A&A...588A..98M, 2016ApJS..225...32B, 2017A&A...598A.100J, 2017AJ....153...21L, 2018A&A...614A..55A, 2018A&A...615A..31D}. The effective temperature and surface gravity consistency between our sample and the spectroscopic comparison sample of 47 stars are quite good: $\langle \Delta T_{\rm eff} \rangle=-11 \pm 58$~K and $\langle \Delta {\rm log}~g \rangle=-0.02 \pm 0.18$. 

\begin{figure}
 \includegraphics[width=\columnwidth]{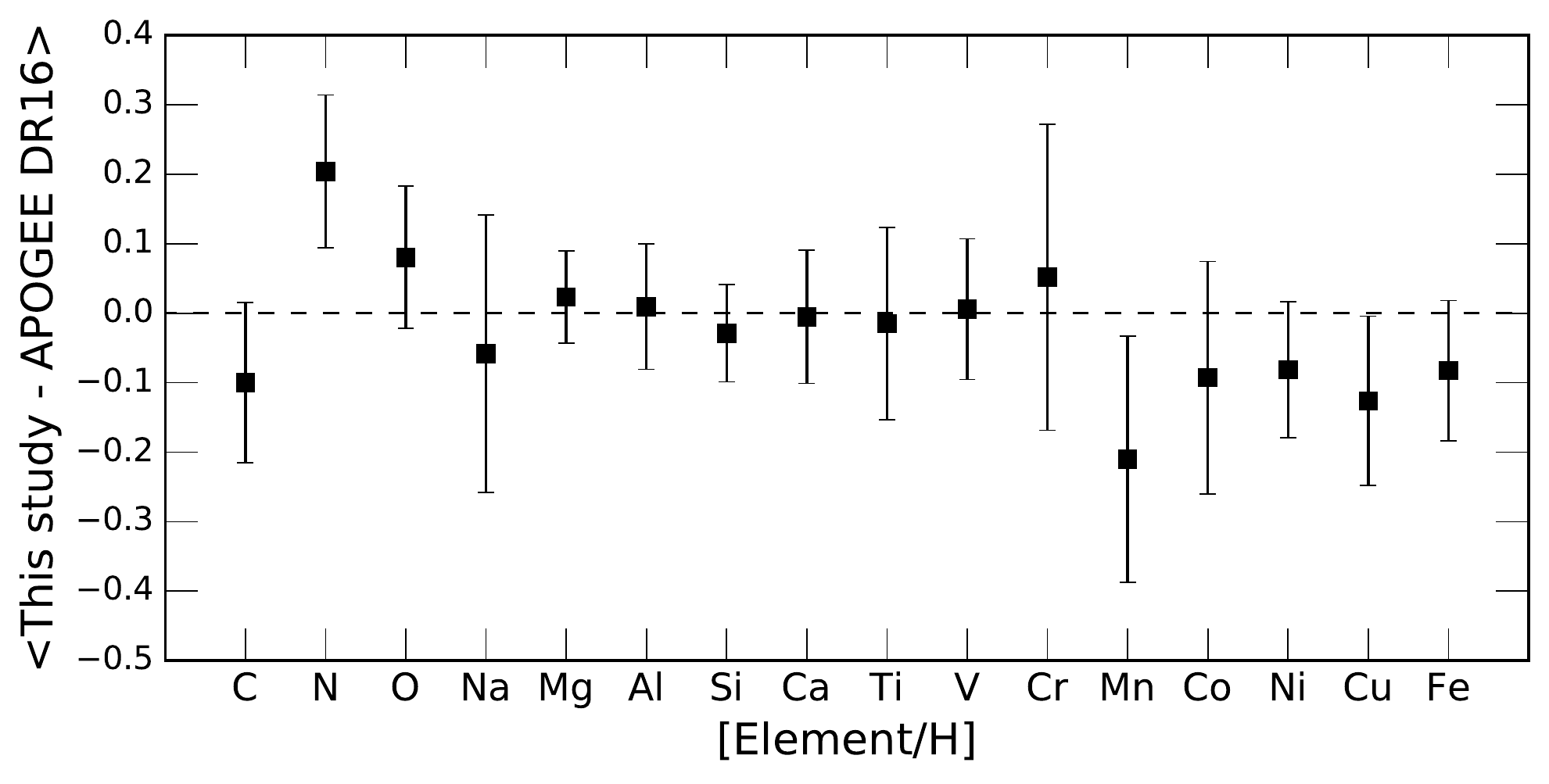}
\caption{Comparison of abundances for up to 18 stars that we have in common with APOGEE DR16.
The average differences and standard deviations are calculated as our values minus the comparison values.}
\label{fig:Oour_Apogee}
\end{figure}

We found 18 stars in common with the near-infrared, large-scale, stellar spectroscopic survey APOGEE 16th data release (DR16) \citep{Ahumada19}. For the comparison, we were using calibrated parameters and abundances determined with the APOGEE Stellar Parameters and Chemical Abundance Pipeline (ASPCAP) \citep{GarciaPerez16}. The biases for the main stellar atmospheric parameters from our sample are $\langle \Delta T_{\rm eff} \rangle=-48 \pm 47$~K and $\langle \Delta {\rm log}~g \rangle=0.06 \pm 0.18$~dex. In Figure~\ref{fig:Oour_Apogee}, we show a comparison of [Element/H] abundances for up to 18 stars in common with the APOGEE DR16.
The average differences for all stars and standard deviations are calculated as our values minus the comparison values. The sample of common stars is not large; however, as one can see, the agreement for the majority of elements is very good. Regarding the C, N, and O elements, the APOGEE survey uses the infrared lines of CH, CN, and OH molecules, respectively, and the agreement for the carbon and oxygen abundances is quite good. The differences of 0.2~dex are between nitrogen and also of manganese abundances. 
As was shown in \citet[][see their Figure 4]{Smith13}, the larger Mn abundances in the APOGEE survey could be caused by a blending of the CN bands. The nitrogen abundance values are accumulating uncertainties of carbon and oxygen abundance determinations apart of others, thus they can be larger than for other chemical elements.

We find a similar $T_{\rm eff}$ bias if we compare our sample with the $Gaia$ DR2 data. $\langle \Delta T_{\rm eff} \rangle=-62 \pm 99$~K for all \noatmospheric~stars. The APOGEE and $Gaia$ DR2 $T_{\rm eff}$ bias is similar probably because the $Gaia$ DR2 $T_{\rm eff}$ data was taken from the Astrophysical Parameters Inference System (Apsis;  \citealt{Andrae2018}) which was trained on a number of samples where the APOGEE giant star input was one of the dominant ones.

\begin{figure}
  \centering
  \includegraphics[width=\columnwidth]{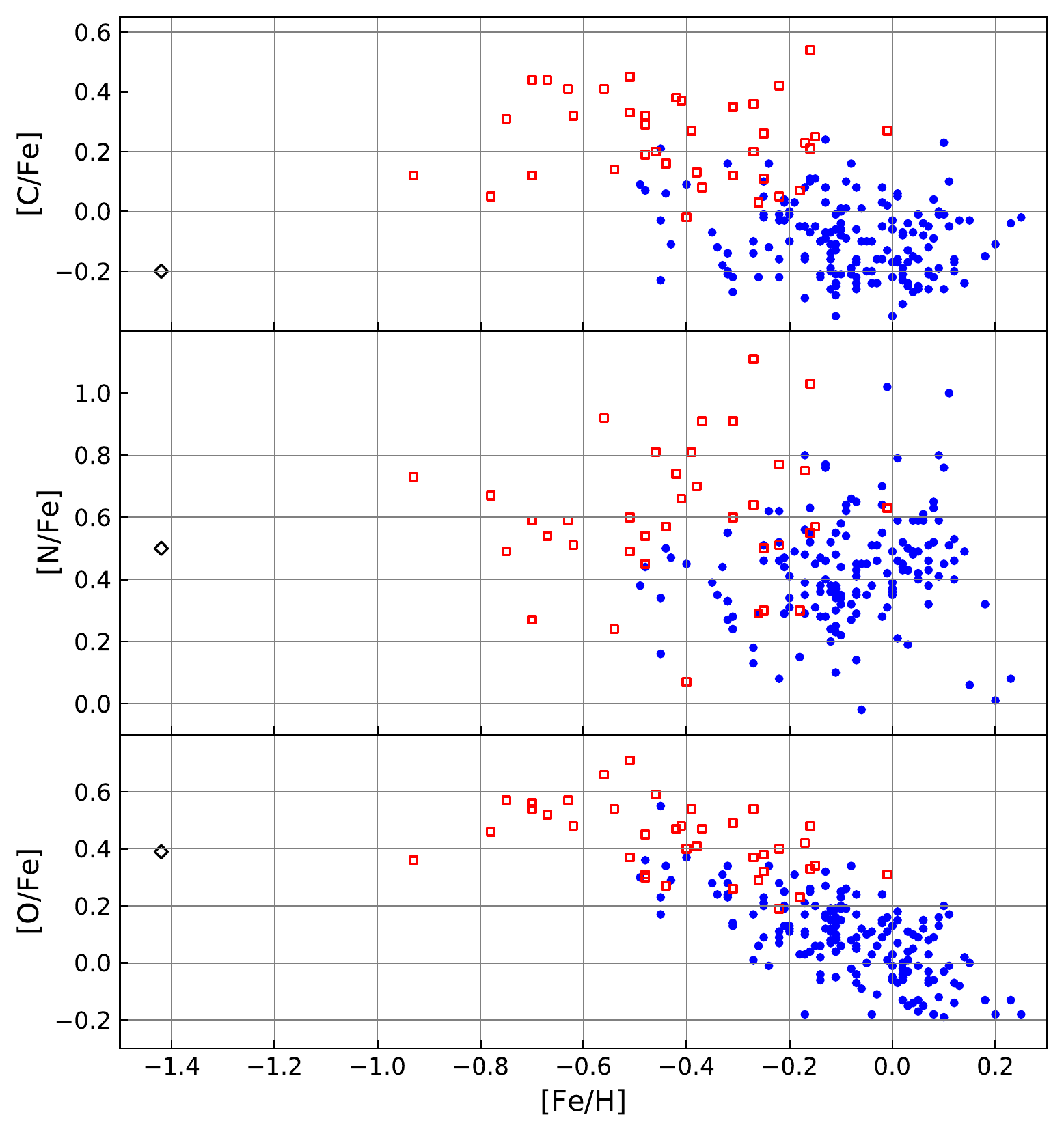}
  \caption{Observed carbon, nitrogen, and oxygen-to-iron element abundance ratios as a function of metallicity. The blue dots, red squares, and the black diamond represent the thin-disk, thick-disk, and halo stars, respectively.}
  \label{fig:CNO}
\end{figure}

\begin{figure*}
\epsscale{0.95}
\plotone{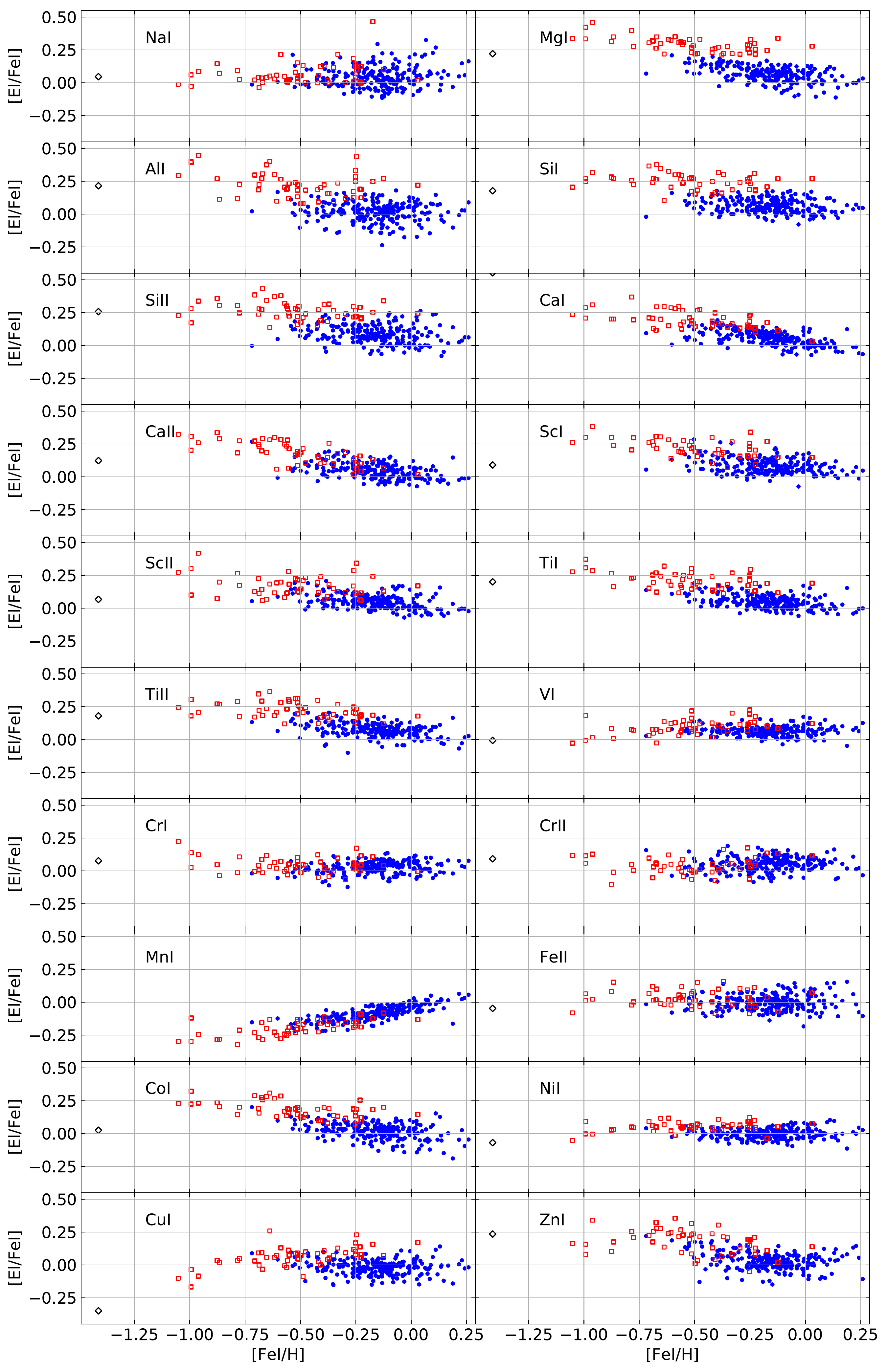}
\caption{[El/\ion{Fe}{1}] ratios as a function of [\ion{Fe}{1}/H]. The meaning of symbols as in Figure~\ref{fig:CNO}. }
\label{fig:allrez}
\end{figure*}

In Table~\ref{tab:CDS}, columns 29--76 contain relative to Solar abundances and uncertainties of C(C$_2$), N(CN), [\ion{O}{1}], \ion{Na}{1}, \ion{Mg}{1}, \ion{Al}{1}, \ion{Si}{1}, \ion{Si}{2}, \ion{Ca}{1}, \ion{Ca}{2}, \ion{Sc}{1}, \ion{Sc}{2}, \ion{Ti}{1}, \ion{Ti}{2}, \ion{V}{1}, \ion{Cr}{1}, \ion{Cr}{2}, \ion{Mn}{1}, \ion{Fe}{1}, \ion{Fe}{2}, \ion{Co}{1}, \ion{Ni}{1}, \ion{Cu}{1}, and  \ion{Zn}{1} for the \noatmospheric ~stars investigated in the 
present study. The abundances are presented in [Element/H] form. The [\ion{Fe}{1}/H] and [\ion{Fe}{2}/H] values occupy 77--80 columns of Table~\ref{tab:CDS}. 
All elemental abundance ratios with respect to [\ion{Fe}{1}/H] are show in  Figures~\ref{fig:CNO} and \ref{fig:allrez} where the stars are colored according to their attribution to the Galactic subcomponents. The stellar attribution to subcomponents is presented in the column 83 of Table~\ref{tab:CDS} and is described in the next section. 

\subsection{Stellar Kinematic Properties, Ages, and Dependence to Galactic Subcomponents}
\label{sec:kinematicparameters}

The majority of the stars have radial velocities between $-40$~and~+20~km\,s$^{-1}$. 
We have compared our radial velocities with the {\it Gaia} DR2 catalog data. The mean and standard deviation of differences between the two sets is $\langle \Delta V_{\rm rad} \rangle=0.05\pm 0.53$~km\,s$^{-1}$. 
Overall radial velocities and corresponding errors are presented in columns 11--12 of Table~\ref{tab:CDS}.

\begin{figure*}
\epsscale{1.15}
\plotone{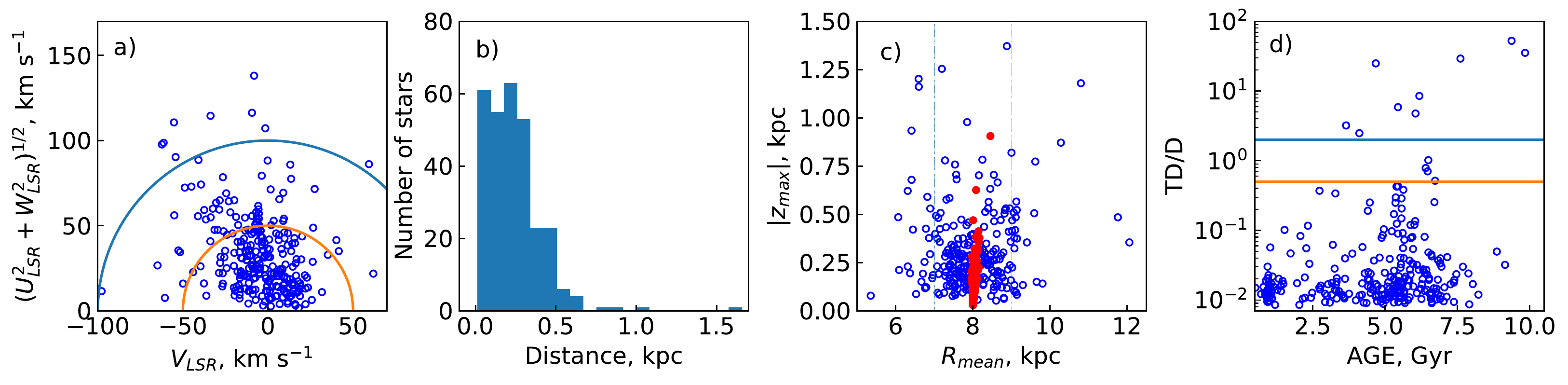}
\caption{Kinematic parameters: (a) Toomre diagram of sample data with lines that show constant values of the total space velocity ($v_{\rm tot}=(U_{\rm LSR}+V_{\rm LSR}+W_{\rm LSR})^{\rm 1/2)}$) at 50 and 100~km\,s$^{\rm -1}$; (b) histogram of distances of the sample stars; (c) distribution of the sample stars in a $z_{\rm max}$ versus $R_{\rm mean}$ plane, where the two vertical dashed lines delimit the Solar neighborhood 7$<R_{\rm gc}<$9~kpc and the red circles are current positions ($|z|$ vs. $R$) for comparison; (d) kinematic thick-to-thin disk probability ratios (TD/D) vs. age for the sample stars, where the upper and lower lines mark TD/D=2.0 and TD/D=0.5, respectively.}
\label{fig:resultsKINEMATICS}
\end{figure*}

\begin{figure*}
  \centering
  \includegraphics[width=0.7\textwidth]{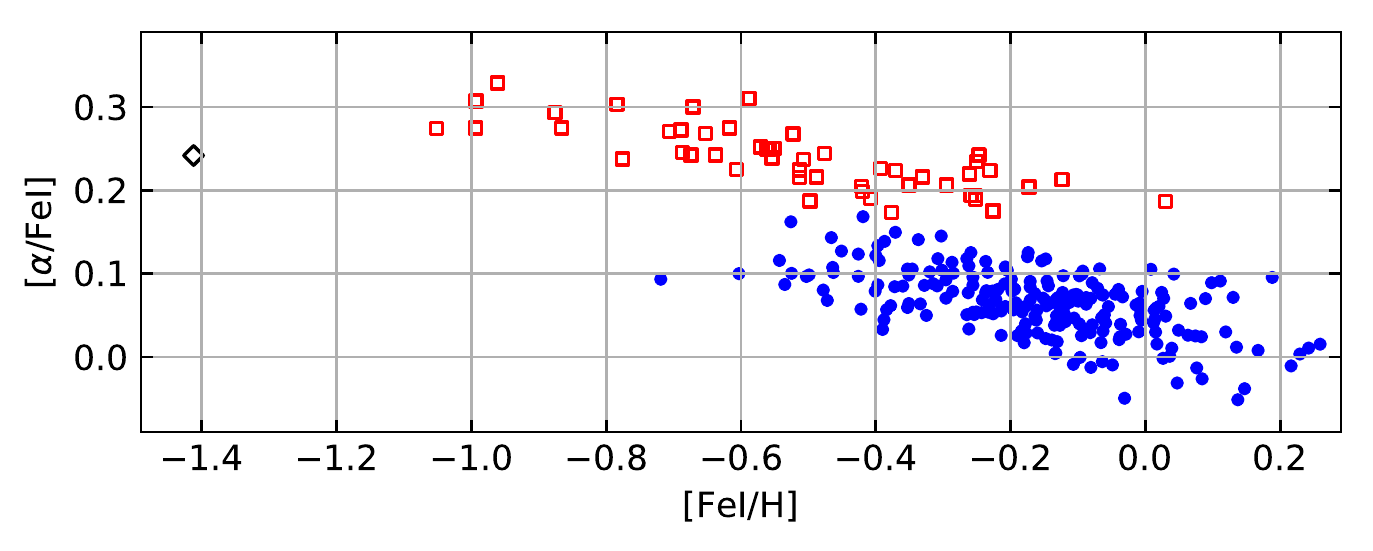}
  \caption{Observed $\alpha$-to-iron element ratios as a function of [\ion{Fe}{1}/H]. [$\alpha$/\ion{Fe}{1}] is an average of \ion{Mg}{1}, \ion{Si}{1}, \ion{Si}{2}, \ion{Ca}{1}, \ion{Ca}{2}, \ion{Ti}{1}, and \ion{Ti}{2}. The meaning of symbols as in Figure~\ref{fig:CNO}.}
  \label{fig:allrez3}
\end{figure*}

The ages of our sample stars are from about 1.0 to 11~Gyr; the majority are close to Solar, about 4~Gyr. 
The age values and uncertainties are presented in columns 13--14 of Table~\ref{tab:CDS}. 

The $U$, $V$, and $W$ velocities; distances, $R_{\rm mean}$, $z_{\rm max}$, and orbital eccentricities, \textit{e}, with corresponding errors are presented in columns 15--27 of Table~\ref{tab:CDS} and are exhibited in Figure~\ref{fig:resultsKINEMATICS}.

It is widely accepted that Galactic subcomponents like thin and thick disks differ in a number of parameters. There are two widely used methods to separate them: kinematical (e.g. \citealt{Bensby2003, Bensby2005, Bensby2014}) and chemical (e.g. \citealt{Adibekyan2012, Recio2014}).

The method introduced by \citet{Bensby2003, Bensby2014} employs the thick-to-thin disk probability ratios. Stars with TD/D~$>$~2 are potential thick-disk stars, stars with TD/D~$<0.5$ potentially belong to the thin disk, and stars with 0.5~$<$~TD/D~$<2.0$ are called "in-between stars". Column 29 in Table~\ref{tab:CDS} presents the thick-to-thin disk probability ratios (TD/D). The panel (d) in Figure~\ref{fig:resultsKINEMATICS} displays the stellar TD/D distribution with age. 
According to the Toomre diagram (panel (d) in Figure~\ref{fig:resultsKINEMATICS}), almost all our stars kinematically belong to the thin or thick disks except HD~175305 which according to \citet{Nissen2010} has a clear halo kinematics ((U$^2_{LSR}$+W$^2_{LSR})^{1/2}$=290~km\,s$^{-1}$, V$_{LSR}=-72$~km\,s$^{-1}$, $R\mathrm{_{mean}}$=16~kpc, and  $z\mathrm{_{max}}$=20.6~kpc). 
Thus, using the kinematical method, we found that our sample consists of 262 thin-disk stars, nine thick-disk stars, and the remaining five stars are "in-between stars" and one halo star.

The chemical separation method can employ [\ion{Mg}{1}/\ion{Fe}{1}] (\citealt{Adibekyan2012}; \citealt{Mikolaitis2014}), [\ion{Ti}{1}/\ion{Fe}{1}] (\citealt{Bensby2014}), or [$\alpha$/\ion{Fe}{1}] (\citealt{Recio2014}) abundance ratios. As in our previous studies, we used [\ion{Mg}{1}/\ion{Fe}{1}] vs. [\ion{Fe}{1}/H] to separate the low-$\alpha$ from high-$\alpha$ stars that potentially belong to  the thin or thick disks, respectively.
The most metal-poor star HD~175305 with halo kinematics should belong to the high-$\alpha$ halo population according to criteria by \citet{Nissen2010}.
Thus from chemical
signatures, we revealed in our sample that there are 219 thindisk,
57 thick-disk, and one high-α halo stars (column 83 in Table~\ref{tab:CDS} presents stars were attributed to the thin or thick disks
or a halo according to the chemical method).

In Figure~\ref{fig:allrez3} we show 
[$\alpha$/\ion{Fe}{1}] which is an average of \ion{Mg}{1}, \ion{Si}{1}, \ion{Si}{2}, \ion{Ca}{1}, \ion{Ca}{2}, \ion{Ti}{1}, and \ion{Ti}{2} (the [$\alpha$/\ion{Fe}{1}] values and standard errors of the mean are presented in columns 81 and 82 of Table~\ref{tab:CDS}). The two disks are separated quite well. Exoplanets found near thin- and thick-disk stars will be of different chemical content.

\subsection{Stars as Potential Planet Hosts}

\begin{figure}
  \includegraphics[width=\columnwidth]{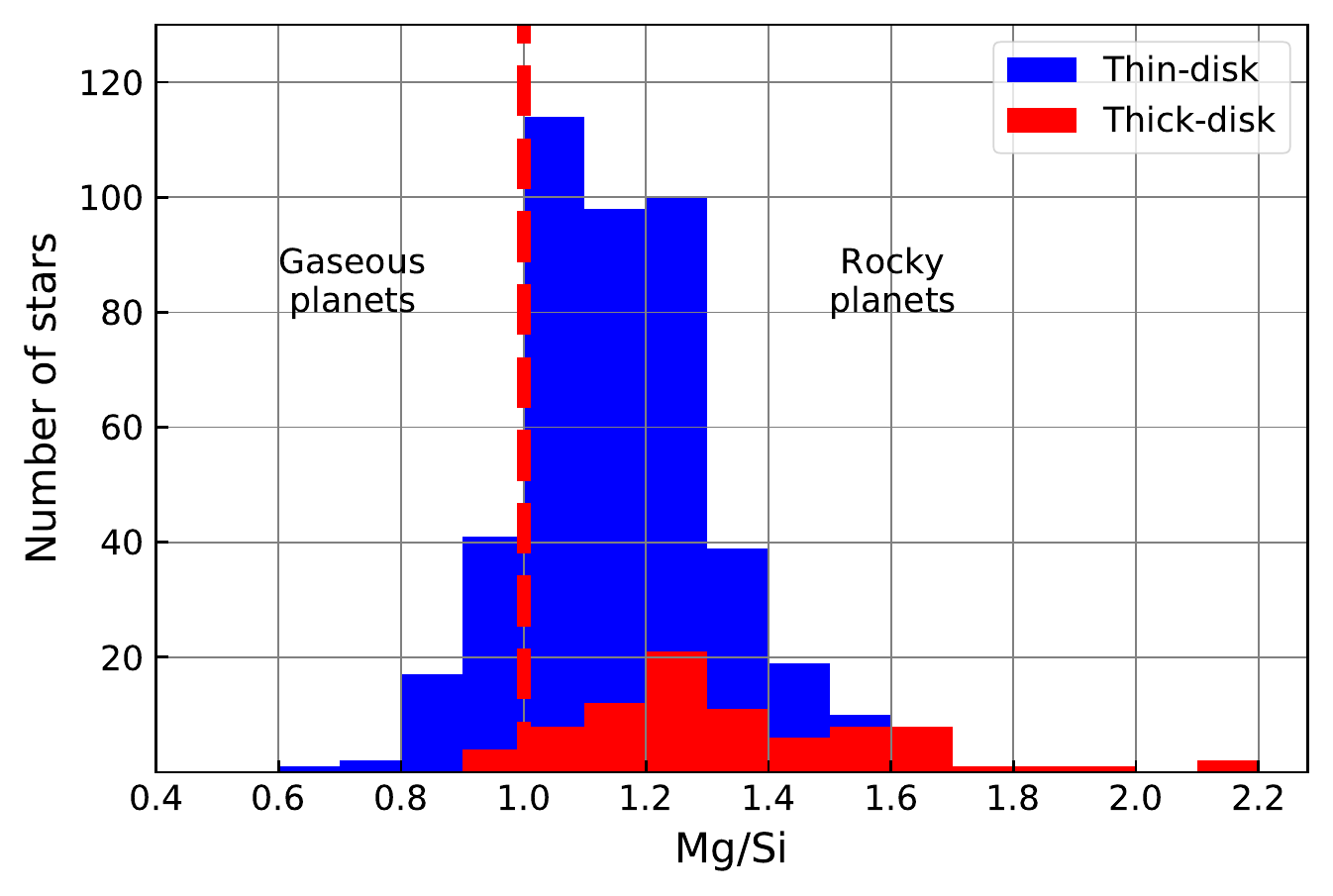}
\caption{Distribution of Mg/Si ratio in thin- and thick-disk stars investigated in this work and \citetalias{Mikolaitis2019}. The vertical dashed line shows the approximate Si/Mg ratio, which separates stars potentially having gaseous and rocky planets, as suggested by e.g. \citealt{Bond10}.}
\label{fig:MgSi}
\end{figure}

The Mg/Si number ratio plays an important role in determining the plate tectonics and habitability of extra-solar planets. 
Theoretical studies have shown that if the Mg/Si number ratio is less than one, the extra-solar planet composition will be mostly made of pyroxene with other silicate based minerals, such as feldspar and a fraction of olivine \citep{Bond10}. Furthermore, if the Mg/Si number ratio is between 1.0 and 2.0, as in our own planet Earth where Mg/Si$_{\odot}$=1.05, the extra-solar planet composition will exist between pyroxene and olivine evenly. And lastly, if the Mg/Si number ratio is more than two, most of the planets will form from material with an olivine structure and the remaining magnesium will form oxides \citep{Bond10, Thiabaud15}. Thus, the planets originating from these three divisions would be different in term of plate tectonics, surface mineral chemistry, and inner geology.

Furthermore, \citet{Dorn15} in their work concluded that in order to constrain the models for interior structure of rocky planets, the stellar chemical abundances, such as iron, silicon, and magnesium, are the key ingredients to reduce degeneracy in interior structure models and to constrain the mantle composition.  

\begin{figure*}
  \centering
  \includegraphics[width=0.99\textwidth]{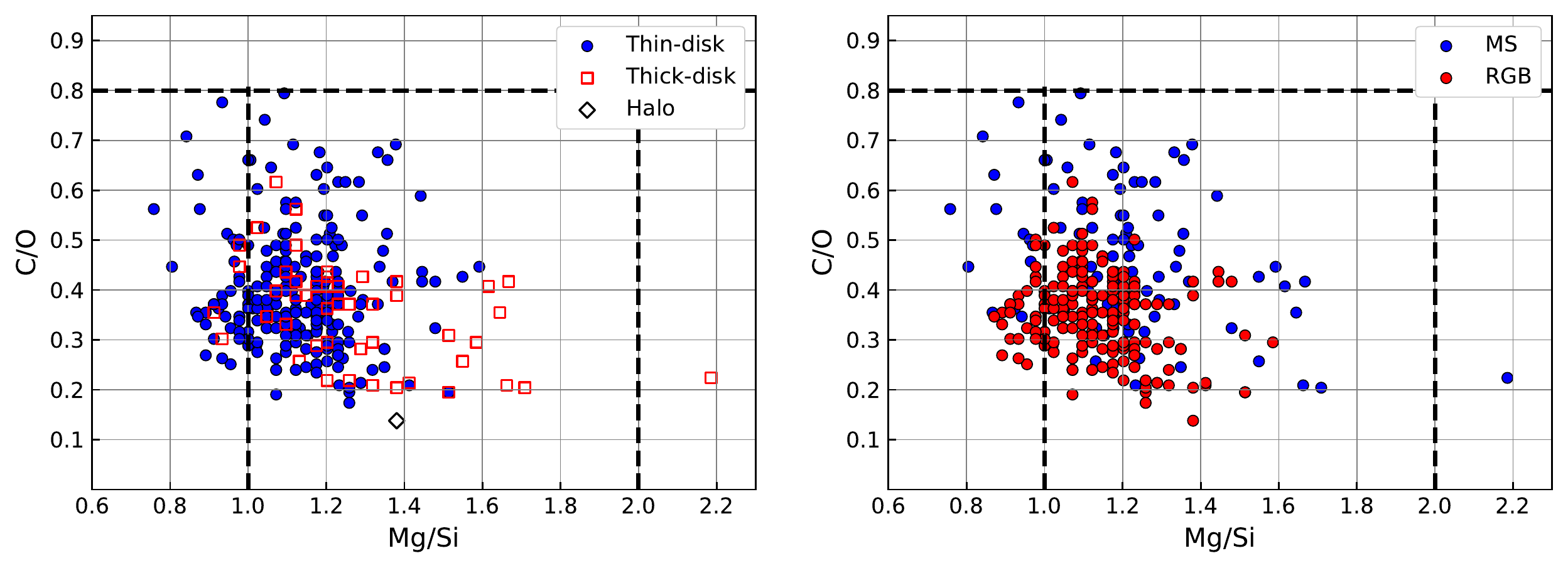}
  \caption{C/O number ratio as a function Mg/Si number ratio. On the left the colors and symbols represent the different Galactic components and on the right the stars are divided into the main-sequence and the red giant branch stars. Together with the data from this work, we plot 249 stars previously investigated with the same instrument and method of analysis (\citetalias{Mikolaitis2019}, \citealt{Stonkute2020}). }
  \label{fig:CO_MGSI}
\end{figure*}

As suggested by, e.g. \citet{Bond10} and \citet{Suarez18}, the number ratio of Mg/Si$=1.0$ marks a division between stars that could form potentially gaseous or rocky planets.  
About 83\,\% of our stars have Mg/Si values in the range between 1.0 and 2.0 with the mean value of Mg/Si=$1.18 \pm 0.13$, which could suggest that they may have planets with a composition close to that of our planet Earth. The remaining $\sim$17\,\% of the stars have Mg/Si~$\leq$~1.0, where planets may have a magnesium-depleted mineralogy. If we add to the sample of 277 stars investigated in this work and 249 stars investigated by \citetalias{Mikolaitis2019} using the same instrument and method of analysis, the percentage of stars in the interval of $1.0<{\rm Mg/Si}<2.0$ increases by 1.4\%. 
\citet{Suarez18} also found that $\sim$83\,\% of their sample of 499 Solar-like stars to have a Mg/Si number ratio between 1.0 and 2.0, while \citet{Brewer16} estimated a somewhat lower percentage among 847 investigated stars: $\sim$64\,\%. The percentage of stars with the particular Mg/Si ratio may vary depending on how many stars belonging to the different Galactic subcomponents are in a sample.  In Figure~\ref{fig:MgSi}, we show a distribution of thin- and thick-disk stars according to their Mg/Si number
ratios for the sample of 526 stars from our and \citetalias{Mikolaitis2019} study. 

In Figure~\ref{fig:CO_MGSI}, we also plotted the stars in the C/O versus Mg/Si number ratio diagrams. The left panel shows where stars are located according their dependence to the Galactic subcomponents, while in the right one, we can see how the C/O ratio differs in dwarfs and giants. Together with the data from this work, we plot 249 stars previously investigated with the same instrument and method of analysis (\citetalias{Mikolaitis2019}, \citealt{Stonkute2020}). It can be seen that in giants, due to evolutionary changes of carbon, the C/O ratio, on average, is lower by about 0.1, thus we have to have in mind that planets that we find around evolved stars were formed when their host stars were young and had larger C/O ratios. 

\section{Summary}

With the aim to contribute in fulfilling the primary goal of the ongoing NASA TESS mission -- to characterize planets around bright and nearby stars -- 
in this paper, we present the main atmospheric parameters, ages, kinematic parameters, and abundances of 24 chemical elements determined from the high-resolution spectroscopy of all bright, ($V<8$~mag), slowly rotating, and cooler than F5 spectral type stars within the northern TESS CVZ.
The observed TESS field also covers the CVZ of JWST, making the region particularly interesting for all types of astronomical studies. 

In the northern TESS CVZ of an $\sim 12$~degree radius, we observed all \noallfield~stars with $V<8$~mag and $(B-V)>0.39$, which roughly corresponds to $T_{\rm eff}$~$<$~6500~K. Among them, 53 stars belong to the JWST CVZ as well. 

 There are 25 stars that appeared to be fast-rotating or double- or even triple-line systems so we decided to postpone for future analyses. A detailed characterization was done for a sample of \noatmospheric~stars of different evolutionary stages, ages, and atmospheric parameters: $T_{\rm eff}$ varied between 3700 and 6600~K and [Fe/H] -- between $-1.5$ and 0.25~dex, ages -- from 1 to 11~Gyr. A distinctive ${\rm log}~g$ distribution clearly separated giant and dwarf stars; the parameter of the former ones varied between 0.8 and 3.5, with a peak at 2.7, while the latter ones displayed values between 3.6 and 4.5 with a peak at 4.3. 
 Data from the {\it Gaia} DR2 catalog was used to calculate stellar kinematic parameters; the mean Galactocentric distances, $R\mathrm{_{mean}}$, span from 5 to 12~kpc and distances from the Galactic plane, $z\mathrm{_{max}}$, reach 1.5~kpc. Stellar velocity components  (\textit{U, V, and W}) were determined as well.

Along with the main atmospheric parameters, abundances of the 24 chemical species determined will serve for the detailed characterization of exoplanets, if discovered around the investigated stars by the NASA TESS space telescope, for the interpretation of exoplanet atmospheres to be made by the upcoming NASA JWST mission and in answering many stellar and Galactic evolution questions. The vast majority of our sample stars ($\sim$83\,\%) exhibit Mg/Si ratios greater than 1.0 and may potentially harbor rocky planets in their systems. 

Only around one third of bright $V<8$~mag stars have spectroscopic observations in the literature. This fact is evident from the star sample of this study as well. Out of  \noatmospheric~stars selected for the spectroscopic analysis, only 47 had previously derived atmospheric parameters from  high-resolution spectroscopy. In the era of exoplanet search among bright stars, this gap of study is more significant than ever, since knowledge of precise stellar atmospheric parameters are very important in characterizing exoplanets and stars themselves. 

\vspace{5mm}

We acknowledge the grant from the European Social Fund via the Lithuanian Science Council (LMTLT) grant No. 09.3.3-LMT-K-712-01-0103. We thank the anonymous referee for a constructive report that helped to improved this paper. We have made extensive use of the NASA ADS and SIMBAD databases.
We are grateful to the Moletai Astronomical Observatory of Vilnius University 
for providing observing time for this project.

\vspace{5mm}

\facility{Exoplanet Archive.}

\software{Astropy \citep{2018AJ....156..123A}, DAOSPEC \citep{Stetson2008}, MOOG \citep{Sneden1973}, galpy \citep{Bovy15},TURBOSPECTRUM \citep{Alvarez1998}, UniDAM \citep{Mints2017}.} 

\appendix
\restartappendixnumbering 
\section{Appendix information}

Table~\ref{tab:CDS} lists the contents of the Machine-readable table (atmopheric parameters, kinematic properties, ages, and individual abundances) together with associated errors, and other information for the investigated stars.

\begin{longtable*}{lllll} 
\caption{Contents of the Machine-readable Table}\\ 
\hline
Col & Label & Units & Explanations \\
\hline
1	&	ID                  	&	  ---   	&	 Tycho catalog identification\\
2	&	TESS\_ID             	&	  ---   	&	 ID in the TESS catalog \\
3	&	Teff                	&	 K    	&	 Effective temperature\\
4	&	eTeff               	&	   K    	&	 Error on effective temperature\\
5	&	Logg                	&	  dex   	&	 Surface gravity\\
6	&	e\_Logg              	&	  dex   	&	 Error on surface gravity\\
7	&	[Fe/H]              	&	  dex   	&	 Metallicity \\
8	&	e\_[Fe/H]            	&	  dex   	&	 Error on metallicity \\
9	&	Vt                  	&	   km s$^{-1}$  	&	 Microturbulence velocity\\
10	&	e\_Vt                	&	  km s$^{-1}$   	&	 Error on microturbulence velocity\\
11	&	Vrad                	&	 km s$^{-1}$ 	&	 Radial velocity\\
12	&	e\_Vrad              	&	 km s$^{-1}$ 	&	 Error on radial velocity \\
13	&	Age                 	&	  log(yr)  	&	 Log age of the star \\
14	&	e\_Age               	&	  log(yr)  	&	 Error on log Age\\
15	&	 U                  	&	  km s$^{-1}$  	&	 $U$ velocity\\
16	&	e\_U                 	&	  km s$^{-1}$   	&	 Error on $U$ velocity\\
17	&	V                   	&	 km s$^{-1}$    	&	 $V$ velocity \\
18	&	e\_V                 	&	 km s$^{-1}$    	&	 Error on $V$ velocity \\
19	&	W                   	&	 km s$^{-1}$    	&	 $W$ velocity \\
20	&	e\_W                 	&	 km s$^{-1}$     	&	 Error on $W$ velocity\\
21	&	 d                  	&	 kpc 	&	 Distance calculated 1/plx\\
22	&	R$_{mean}$           	&	 kpc    	&	 Mean Galactocentric distance\\
23	&	e\_R$_{mean}$        	&	 kpc 	&	Error on mean Galactrocentric distance\\
24	&	z$_{max}$                	&	 kpc    	&	 Distance from Galactic plane\\
25	&	e\_z$_{max}$              	&	 kpc    	&	 Error on distance from Galactic plane\\
26	&	{\it{e}}                   	&	 ---    	&	 Orbital eccentricity\\
27	&	e\_{\it{e}}                 	&	 ---  	&	 Error on orbital eccentricity\\
28	&	TD/D                	&	 --- 	&	 Thick-to-thin disk probability ratio\\
29	&	[C/H]    	&	 dex    	&	 Carbon abundance \\
30	&	e\_[C/H]  	&	 dex     	&	 Error on carbon abundance \\
... \\
75	&	[\ion{Zn}{1}/H]    	&	 dex    	&	 Zinc abundance \\
76	&	e\_[\ion{Zn}{1}/H]  	&	 dex     	&	 Error on zinc abundance\\
77	&	[\ion{Fe}{1}/H]    	&	 dex    	&	  Iron abundance\\
78	&	e\_\ion{Fe}{1}/H]  	&	 dex    	&	 Error on iron abundance\\
79	&	[\ion{Fe}{2}/H]    	&	 dex    	&	 Ionized iron abundance \\
80	&	e\_[\ion{Fe}{2}/H]  	&	 dex     	&	 Error on ionized iron abundance\\
81  & [{alpha}/\ion{Fe}{1}] & dex &  Averaged \ion{Mg}{1}, \ion{Si}{1}, \ion{Si}{2}, \ion{Ca}{1}, \ion{Ca}{2}, \ion{Ti}{1}, and \ion{Ti}{2} to \ion{Fe}{1} abundance ratio \\
82  & e\_[{alpha}/\ion{Fe}{1}] & dex &  Standard error of the mean on [{alpha}/\ion{Fe}{1}] \\
83	&	Group  	&	 ---     	&	 Chemical attribution to the Galactic subcomponent\\
\noalign{\smallskip}
\hline
\label{tab:CDS}

\end{longtable*}


\bibliography{TESS}{}
\bibliographystyle{aasjournal}

\end{document}